\documentclass[sigconf,nonacm]{acmart}  

\usepackage{amssymb}

\renewcommand\footnotetextcopyrightpermission[1]{}

\AtBeginDocument{%
  \providecommand\BibTeX{{%
    \normalfont B\kern-0.5em{\scshape i\kern-0.25em b}\kern-0.8em\TeX}}}

\copyrightyear{2025} 
\acmYear{2025} 
\setcopyright{acmcopyright}\acmConference[SIGSPATIAL]{The 5th ACM SIGSPATIAL International Workshop on AI for Geographic Knowledge Discovery (GeoAI '22)}{November 3, 2025}{Minneapolis, MN, USA}
\acmBooktitle{The 33th ACM SIGSPATIAL, November 3, 2025, Minneapolis, MN, USA}
\acmPrice{}
\acmDOI{}
\acmISBN{}

\begin{document}

\title{Reframing Spatial Dependence as Geographic Feature Attribution}

\author{Chuan Chen}
\email{chuan.chen@tum.de}
\affiliation{%
  \institution{Technical University of Munich}
  \streetaddress{ Arcisstraße 21}
  \city{Munich}
  \state{Bavaria}
  \country{Germany}
  \postcode{80333}
}

\author{Peng Luo}
\authornote{Contact author}
\email{pengluo@mit.edu}
\affiliation{%
  \institution{Massachusetts Institute of Technology}
  \city{Cambridge}
  \country{USA}}

\renewcommand{\shortauthors}{Chen and Luo.}

\begin{abstract}

Spatial dependence—referring to the correlation between variable values observed at different geographic locations—is one of the most fundamental characteristics of spatial data. Identifying and quantifying such dependence is critical for deepening our understanding of spatial phenomena. The presence of spatial dependence violates the classical statistical assumption of independent and identically distributed (i.i.d.) observations and implies a high degree of information redundancy within spatial datasets. However, this redundancy can also be interpreted as structured information, which has been widely leveraged in spatial modeling, prediction, and explanation tasks.
With the rise of geospatial big data and the rapid advancement of deep learning and large models, effectively modeling and characterizing spatial dependence has become essential for enhancing the performance of spatial analysis and uncovering latent spatial processes. From a data-driven perspective, this study proposes a novel interpretation: spatial dependence can be understood as the contribution of geographic location—specifically, latitude and longitude—to the observed variation in target variables.
To validate this hypothesis, we conduct a series of simulation experiments in which data are generated based on known spatial processes. We train machine learning models to predict variable values using only coordinate information. Subsequently, interpretable machine learning techniques are employed to quantify the contribution of spatial features. The resulting importance scores are then compared with local indicators of spatial association (LISA). Across a range of spatial process settings, we observe consistently high correlations (greater than 0.94) between coordinate-based contributions and LISA values. We further validate this assumption in real-world datasets- housing prices in California.
These findings offer a new data-driven perspective on spatial dependence, bridging traditional spatial statistical approaches with modern machine learning techniques.

\end{abstract}





\maketitle

\section{INTRODUCTION}

A key challenge in geospatial artificial intelligence (GeoAI) lies in accurately modeling and representing the unique characteristics of geographic data \citep{luo2024generalized,luo2025understanding}. Among these characteristics, spatial dependence is one of the most prominent \citep{anselin1990spatial}. Effectively capturing spatial dependence is essential for understanding the patterns and processes behind geographic variables and for guiding the construction of spatial models.

There are two main approaches to modeling spatial dependence. The first focuses on developing statistical indicators, such as Moran’s I and Local Indicators of Spatial Association (LISA), which are used to measure the strength of spatial dependence in geographic data \citep{anselin1995local}. However, these metrics are descriptive and are not designed for predictive tasks. The second approach assumes the existence of spatial dependence and incorporates this assumption into predictive models to infer unobserved values. One of the most common strategies is the distance decay principle, which has been widely applied in building the spatial weight matrix to measure spatial interactions between locations \citep{fotheringham2009geographically,soininen2007distance}.

However, both approaches struggle to offer a general and unified understanding of spatial dependence. In the era of geospatial big data and large models, proposing a data-driven interpretation of spatial dependence can bridge the gap between geographic data and computer science. This perspective may enhance deep learning models' ability to understand and model geographic phenomena. Moreover, spatial dependence enables models to extract more information from limited observations, which may improve computational efficiency and support scalable GeoAI modeling.

In this study, we hypothesize that spatial dependence, from a data-driven perspective, can be interpreted as the predictive power of geographic coordinates (e.g., latitude and longitude) on the distribution of a target variable. Intuitively, if a variable exhibits strong spatial autocorrelation, its values tend to be spatially clustered rather than randomly distributed. In such cases, geographic location itself contains substantial information about the variable’s magnitude, implying that the variable can be predicted to a large extent using only spatial coordinates.

To test this hypothesis, we conduct a set of simulation experiments. For each simulated dataset generated under different spatial processes, we first compute spatial autocorrelation metrics (LISA). Then, we train machine learning models to predict the target variable using only geographic coordinates and use SHAP values to explain the contribution of location features. Finally, we examine the relationship between spatial autocorrelation and the contribution of geographic features. We further validate this hypothesis using a real-world dataset—housing prices in California.

\section{Methodology}

We propose a coordinate-based interpretation of spatial dependence by comparing classical spatial statistics with explainable machine learning. Specifically, we define SHAP\_Geo as the combined SHAP values for latitude and longitude obtained from models trained solely on geographic coordinates. This reflects the model-learned spatial attribution at each location.

To validate our hypothesis that spatial dependence can be reframed as coordinate-based explanatory power, we compare SHAP\_Geo against Local Indicators of Spatial Association (LISA), a standard spatial autocorrelation measure. The comparison involves computing the Pearson correlation between the absolute SHAP\_Geo and LISA scores across spatial units.

Our experimental design involves:
(1) generating synthetic datasets with known spatial patterns;
(2) training coordinate-only models to predict spatial target values;
(3) computing SHAP\_Geo and LISA for each point;
(4) assessing the correspondence between them via correlation analysis;
(5) applying the same framework to real-world data.

This streamlined approach enables a direct, interpretable comparison between model-derived spatial attributions and traditional spatial structure.

\subsection{Simulation Setup and Notation}

We consider a geospatial regression task with $n$ spatial samples, each having coordinates $(x_i, y_i)$ and a target value $z_i$. Spatial dependence is assessed using two approaches: traditional LISA scores and coordinate-based SHAP attribution, denoted as SHAP\_Geo.

The key variables and notation are summarized in Table~\ref{tab:variables}.

\begin{table}[htbp]
\centering
\small
\caption{Key variables for spatial setup}
\label{tab:variables}
\begin{tabular}{lp{6cm}}
\toprule
Symbol & Definition \\
\midrule
$n$ & Number of spatial samples \\
$(x_i, y_i)$ & Coordinates of point $i$ \\
$z_i$ & Observed target value \\
$\bar{z}$ & Mean of $z_i$ \\
$S^2$ & Variance of $z_i$ \\
$w_{ij}$ & Spatial weight ($w_{ii} = 0$) \\
$k$ & Constant weight: $w_{ij} = k$ for $i \neq j$ \\
$f(x_i, y_i)$ & Coordinate-based prediction model \\
$\hat{z}_i$ & Predicted value at $i$ \\
$\mu_{\hat{z}}$ & Mean of $\hat{z}_i$ \\
$\phi_{x,i}, \phi_{y,i}$ & SHAP values for $x_i$, $y_i$ \\
SHAP\_Geo$_i$ & $\phi_{x,i} + \phi_{y,i}$ \\
\bottomrule
\end{tabular}
\end{table}

This formulation enables direct comparison between SHAP\_Geo and LISA, revealing how much spatial variation can be explained solely by geographic location.

\subsection{LISA Coefficient and SHAP\_Geo Attribution}

We compare spatial autocorrelation and coordinate-based attribution using two measures: the local Moran's $I_i$ and SHAP\_Geo. Local Moran's $I_i$ quantifies spatial clustering as:

\begin{equation}
I_i = \frac{(z_i - \bar{z}) \sum_{j \neq i} w_{ij}(z_j - \bar{z})}{S^2},
\end{equation}

where $\bar{z}$ is the sample mean and $S^2$ is the variance of $z_i$. Under uniform weights ($w_{ij} = k$), $I_i$ simplifies to:

\begin{equation}
I_i = -k \cdot \frac{(z_i - \bar{z})^2}{S^2}.
\end{equation}

This captures each location’s deviation from the global mean.

We define SHAP\_Geo as the sum of SHAP values for coordinates $(x_i, y_i)$ from a model trained only on spatial inputs:

\begin{equation}
\text{SHAP\_Geo}_i = \phi_{x,i} + \phi_{y,i} = \hat{z}_i - \mu_{\hat{z}},
\end{equation}

where $\hat{z}_i$ is the model prediction and $\mu_{\hat{z}}$ is the mean prediction. SHAP\_Geo reflects the learned spatial effect at point $i$.

\subsection{Comparing SHAP\_Geo with LISA}

To assess alignment between spatial autocorrelation and coordinate-based attribution, we compute the Pearson correlation $\rho$ between $I_i$ and $|\text{SHAP\_Geo}_i|$:

\begin{equation}
\rho = \frac{ \sum_{i}(I_i - \bar{I})(|\text{SHAP\_Geo}_i| - \overline{|\text{SHAP\_Geo}|}) }
             { \sqrt{ \sum_{i}(I_i - \bar{I})^2 } \cdot \sqrt{ \sum_{i}(|\text{SHAP\_Geo}_i| - \overline{|\text{SHAP\_Geo}|})^2 } }.
\end{equation}

A high $\rho$ indicates that model-based spatial attribution aligns with traditional measures of spatial dependence.

\subsection{Experimental Setup}

\textit{Synthetic experiment:} We generate $n$ points with spatial coordinates $(x_i, y_i)$ and simulate spatial processes (e.g., linear gradients, radial patterns). A regression model $f(x_i, y_i)$ predicts $z_i$, from which SHAP\_Geo is computed. LISA is calculated from $z_i$, and we analyze correlation between the two.

\textit{Real-world experiment:} We apply the same process to the California housing dataset, using median house value as the target. We compute SHAP\_Geo from a coordinate-only model and compare it to LISA derived from observed prices.

To capture spatial co-dependence, we also compute the bivariate Moran’s $I$:

\begin{equation}
I_{XY} = \frac{ \sum_{i} (x_i - \bar{x}) \sum_{j} w_{ij} (y_j - \bar{y}) }
              { \sqrt{ \sum_{i} (x_i - \bar{x})^2 } \cdot \sqrt{ \sum_{j} (y_j - \bar{y})^2 } },
\end{equation}

where $x_i = I_i$ and $y_i = \text{SHAP\_Geo}_i$. This measures spatial alignment between the two surfaces beyond pointwise correlation.

\section{Case study}

\subsection{Data and Experiment}

\subsubsection{Synthetic experiment}

To evaluate the relationship between spatial autocorrelation and SHAP\_Geo, we construct synthetic datasets over a $100 \times 100$ grid within a bounding box (latitude $[10, 50]$, longitude $[-120, -70]$), yielding $n = 10{,}000$ points. Each spatial pattern is designed to simulate structured variation (e.g., linear gradient, radial divergence).

For each pattern:
(1) The dataset is split into training/testing sets (80/20);
(2) LISA scores $I_i$ are computed using 8-nearest-neighbor weights;
(3) An XGBoost model is trained with coordinates $(x_i, y_i)$ to predict values $A_i$;
(4) SHAP values are summed to obtain $\text{SHAP\_Geo}_i = \phi_{x,i} + \phi_{y,i}$;
(5) We compute Pearson correlation between $I_i$ and $\text{SHAP\_Geo}_i$.

This setup enables controlled analysis of how well coordinate-based attribution reflects spatial structure.

\subsubsection{Real-World Experiment: California Housing}

We apply the same framework to the California housing dataset, which includes $n = 20{,}640$ block group records with coordinates and socio-economic attributes. We focus on seven variables, \textit{MedInc}, \textit{HouseAge}, \textit{AveRooms}, \textit{AveBedrms}, \textit{Population}, \textit{AveOccup}, and \textit{MedHouseValue}.

For each variable:
(1) We train a coordinate-only model and compute SHAP\_Geo;
(2) Local Moran’s $I^{(A)}_i$ is computed from observed values;
(3) Pearson correlation is computed between $|\text{SHAP\_Geo}_i|$ and $I^{(A)}_i$;
(4) Bivariate Moran’s $I$ is used to assess spatial cross-dependence.

These real-world experiments validate that SHAP\_Geo captures interpretable spatial structure in both synthetic and observed geographic data.

\subsection{Results}

\subsubsection{Synthetic Spatial Patterns}

\begin{figure}[H]
    \centering
    \includegraphics[width=\linewidth]{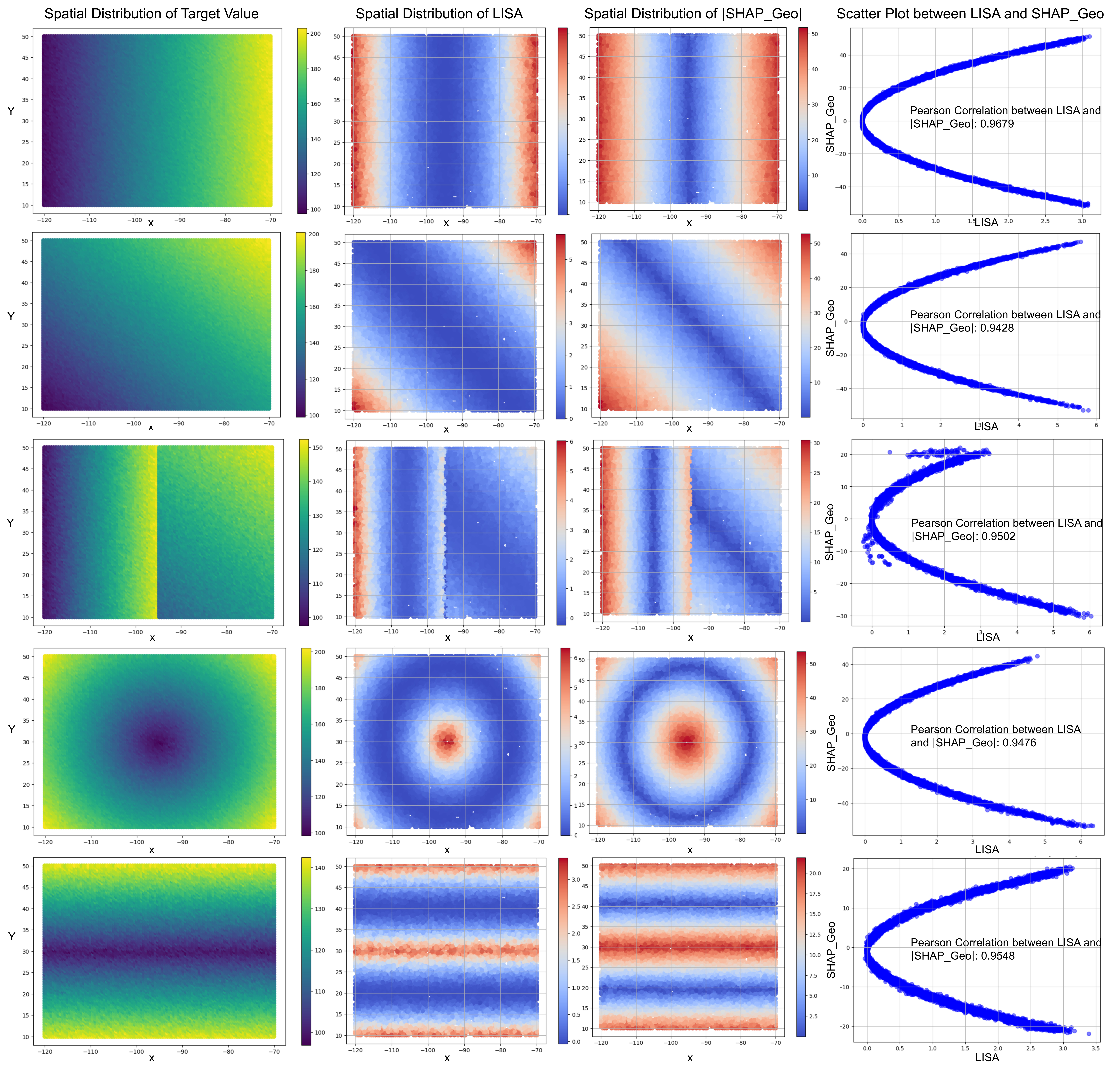}
    \caption{Visual comparison of LISA and SHAP\_Geo across five synthetic spatial patterns. From left to right: (1) target value $A$, (2) LISA $I_i$, (3) $|\text{SHAP\_Geo}_i|$, (4) scatter plot and correlation.}
    \label{fig:synthetic_results}
\end{figure}

Figure~\ref{fig:synthetic_results} shows results across five synthetic patterns. Each pattern displays: (1) the target variable $A$, (2) local Moran’s $I_i$, (3) $|\text{SHAP\_Geo}_i|$, and (4) a scatter plot with the Pearson correlation $\rho$.

Across all cases, SHAP\_Geo closely mirrors LISA in spatial structure and intensity. The Pearson correlations are consistently above 0.94, with highest alignment in the horizontal ($\rho = 0.9679$), V-shaped ($\rho = 0.9548$), and radial ($\rho = 0.9476$) patterns. Even in mixed and diagonal settings, SHAP\_Geo effectively captures directional variation. These results validate SHAP\_Geo’s ability to represent spatial dependence through model-learned coordinate attribution.

\subsubsection{Real-World: California Housing}

\begin{figure}[H]
    \centering
    \includegraphics[width=\linewidth]{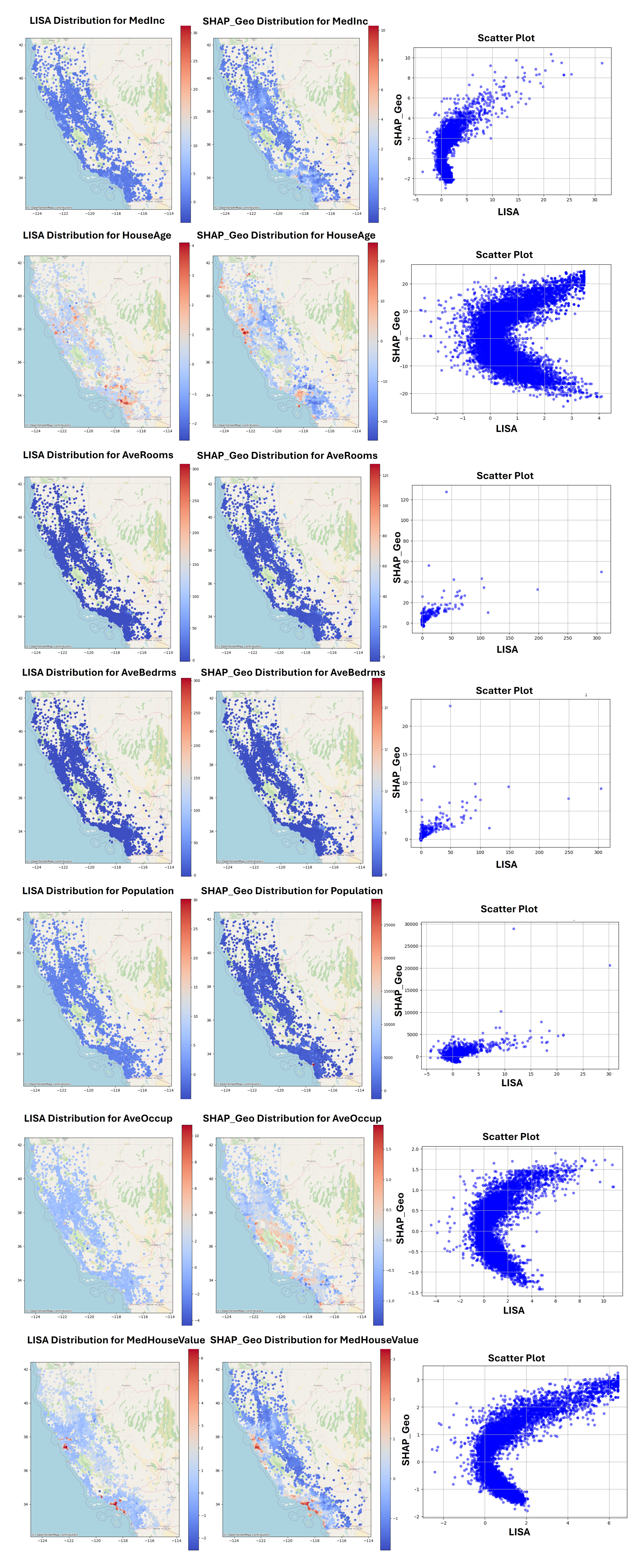}
    \caption{Comparison of LISA and $|\text{SHAP\_Geo}|$ for seven variables in the California housing dataset. Each row: LISA (left), SHAP-based attribution (middle), and correlation plot (right).}
    \label{fig:realworld_results}
\end{figure}

Figure~\ref{fig:realworld_results} presents results on seven variables from the California housing dataset. SHAP\_Geo and LISA show strong spatial alignment for variables like \texttt{MedHouseValue}, with more diffuse but interpretable patterns in others like \texttt{Population}.

Table~\ref{tab:pearson_moran_results} summarizes the correlations. Pearson values range from 0.64 to 0.88, and bivariate Moran’s $I$ from 0.34 to 0.86, confirming that SHAP\_Geo reliably tracks classical spatial structure. The strongest alignment appears for \texttt{MedHouseValue} ($\rho = 0.8781$, $I = 0.8604$), supporting the validity of coordinate-based SHAP attribution in real-world applications.

\begin{table}[H]
\centering
\caption{Pearson and Bivariate Moran’s I between LISA and $|\text{SHAP\_Geo}|$}
\label{tab:pearson_moran_results}
\begin{tabular}{lcc}
\toprule
\textbf{Variable} & \textbf{Pearson} & \textbf{Bivariate Moran’s I} \\
\midrule
MedInc & 0.7806 & 0.6607 \\
HouseAge & 0.7717 & 0.7868 \\
AveRooms & 0.6557 & 0.4030 \\
AveBedrms & 0.6982 & 0.4068 \\
Population & 0.6359 & 0.3406 \\
AveOccup & 0.7795 & 0.7817 \\
MedHouseValue & 0.8781 & 0.8604 \\
\bottomrule
\end{tabular}
\end{table}

Overall, these results demonstrate that SHAP\_Geo—computed solely from model-based attributions of spatial coordinates—can reveal spatial structures that are statistically aligned with classical measures of spatial autocorrelation.

\section{CONCLUSION}

In this study, we explore spatial dependence from a novel explainable machine learning perspective. Specifically, we propose SHAP\_Geo, a coordinate-based attribution method derived from SHAP values, to quantify the spatial contribution learned by predictive models. We compare SHAP\_Geo against classical spatial statistics, namely local Moran's I, on both synthetic and real-world datasets. Experiments on controlled spatial patterns show that SHAP\_Geo and 
LISA exhibit highly similar spatial distributions, with Pearson correlations consistently above 0.94. On real-world California housing data, we observe a strong alignment between SHAP\_Geo and LISA across seven attributes, further supported by bivariate Moran’s I scores. These results suggest that SHAP-based coordinate attributions can reflect spatial structure without explicitly modeling spatial dependence. As SHAP\_Geo requires only model predictions and coordinate inputs, it offers a lightweight yet effective diagnostic tool for spatial modeling tasks. Future work will focus on extending SHAP\_Geo to incorporate temporal dynamics and causal spatial interactions, as well as integrating it with spatial clustering and regime detection frameworks to better support responsible and interpretable geospatial AI.

\bibliographystyle{ACM-Reference-Format}
\bibliography{Reference}

\end{document}